\documentclass[conference]{IEEEtran}
\usepackage{cite}
\usepackage{amsmath,amssymb,amsfonts}
\usepackage{textcomp}
\usepackage{hyperref} 
\usepackage{setspace} 

\usepackage{balance} % for balancing columns on the final page
\usepackage{times}
\usepackage{soul}
\usepackage{url}
\usepackage[utf8]{inputenc}
\usepackage{graphicx}
\usepackage{amsthm}
\usepackage{booktabs}
\usepackage[ruled, vlined, linesnumbered]{algorithm2e}
\usepackage{xcolor}
\usepackage{soul}
\usepackage[small]{caption}
\usepackage{url}
\let\savedegree\bigtimes
\let\bigtimes\relax
\usepackage{mathabx}
\let\bigtimes\savedegree
\usepackage{mathabx}
\usepackage{multirow}
\usepackage{cases}
\usepackage[noend]{algpseudocode}
\usepackage{verbatim}
\usepackage{enumitem}
\usepackage{mathrsfs}
\usepackage{tikz}
\newtheorem{theorem}{Theorem}

\theoremstyle{plain}

\makeatletter
\newcommand{\linebreakand}{%
  \end{@IEEEauthorhalign}
  \hfill\mbox{}\par
  \mbox{}\hfill\begin{@IEEEauthorhalign}
}
\makeatother

\makeatletter \def\@IEEEpubidpullup{8\baselineskip} \makeatother

\def\BibTeX{{\rm B\kern-.05em{\sc i\kern-.025em b}\kern-.08em
    T\kern-.1667em\lower.7ex\hbox{E}\kern-.125emX}}

\usepackage{fancyhdr}
\usepackage{kantlipsum}
\fancypagestyle{plain}{
\fancyhf{}
\fancyhead[C]{Conference on \LaTeX} %% C or L or R. %\fancyfoot[L]{This is a notice}% %% C or L or R. \renewcommand{\footrulewidth}{0pt} %\renewcommand{\headrulewidth}{0pt} 
} \usepackage{eso-pic}

\begin{document}

\IEEEoverridecommandlockouts
\IEEEpubid{
\parbox{\columnwidth}{\vspace{-4\baselineskip} Permission to make digital or hard copies of all or part of this work for personal or classroom use is granted without fee provided that copies are not made or distributed for profit or commercial advantage and that copies bear this notice and the full citation on the first page. Copyrights for components of this work owned by others than the author(s) must be honored. Abstracting with credit is permitted. To copy otherwise, or republish, to post on servers or to redistribute to lists, requires prior specific permission and/or a fee. Request permissions from \href{mailto:permissions@acm.org}{permissions@acm.org}.\hfill\vspace{-0.8\baselineskip}\\ \begin{spacing}{1.2}
\small\textit{ASONAM '23}, November 6-9, 2023, Kusadasi, Turkey \\
\copyright\space2023 Association for Computing Machinery. \\
ACM ISBN 979-8-4007-0409-3/23/11 \$15.00 \\ \url{https://doi.org/10.1145/3625007.3627316}
\end{spacing}
\hfill}
\hspace{0.9\columnsep}\makebox[\columnwidth]{\hfill}}
\IEEEpubidadjcol 

\AddToShipoutPictureBG*{
\AtPageUpperLeft{
\setlength\unitlength{1in}
\hspace*{\dimexpr0.5\paperwidth\relax}%% change \dimexpr0.5\paperwidth\relax appropriately 
\makebox(0,-0.75)[c]{\textbf{2023 IEEE/ACM International Conference on Advances in Social Networks Analysis and Mining (ASONAM)}}}}

\title{Efficient size-prescribed $k$-core search
}

\author{
\IEEEauthorblockN{ Yiping Liu\textsuperscript{1} \\ 
yliu823@aucklanduni.ac.nz
}
\and
\IEEEauthorblockN{ Bo Yan\textsuperscript{2}
\\ yanbo@bit.edu.cn
}
\linebreakand 
\IEEEauthorblockN{ Bo Zhao\textsuperscript{2}
\\ xbackturbo@163.com
}
\and 
\IEEEauthorblockN{ Hongyi Su\textsuperscript{2}
\\ henrysu@bit.edu.cn
}
\and 
\IEEEauthorblockN{ Yang Chen\textsuperscript{1}
\\ yang.chen@auckland.ac.nz
}
\and 
\IEEEauthorblockN{ Michael Witbrock\textsuperscript{1}
\\ m.witbrock@auckland.ac.nz
}
\linebreakand
\IEEEauthorblockA{1
\textit{School of computer science,Univsersity of Auckland, Auckland, New Zealand}\\
2 \textit{School of Computer Science, Beijing Institute of Technology, BeiJing, China}
}
}

\maketitle

\begin{abstract}
$k$-core is a subgraph where every node has at least $k$ neighbors within the subgraph.
The $k$-core subgraphs has been employed in large platforms like Network Repository to comprehend the underlying structures and dynamics of the network.
Existing studies have primarily focused on finding $k$-core groups without considering their size, despite the relevance of solution sizes in many real-world scenarios. 
This paper addresses this gap by introducing the size-prescribed $k$-core search (SPCS) problem, where the goal is to find a subgraph of a specified size that has the highest possible core number.
We propose two algorithms, namely the {\it TSizeKcore-BU} and the {\it TSizeKcore-TD}, to identify cohesive subgraphs that satisfy both the $k$-core requirement and the size constraint. 
Our experimental results demonstrate the superiority of our approach in terms of solution quality and efficiency. 
The {\it TSizeKcore-BU} algorithm proves to be highly efficient in finding size-prescribed $k$-core subgraphs on large datasets, making it a favorable choice for such scenarios. On the other hand, the {\it TSizeKcore-TD} algorithm is better suited for small datasets where running time is less critical.
\end{abstract}

\begin{IEEEkeywords}
Social network, $k$-core, community detection, subgraph search, prescribed size.
\end{IEEEkeywords}

\section{Introduction}
Graphs are widely used to model relationships between individuals with wide applications in various domains, such as social science, biology, information technology, and physics. Within graph analysis, a fundamental challenge lies in the discovery of $k$-cores, which are subgraphs where every node has at least $k$ neighbors within the subgraph. These $k$-core groups play a significant role in comprehending the underlying structures and dynamics of the network \cite{west2001introduction}.

Despite the extensive research on $k$-core group detection, existing studies primarily focus on finding $k$-core groups without considering their size, neglecting an important aspect of network analysis. However, in many real-world scenarios, researchers and practitioners are interested in identifying $k$-core groups with a specific and prescribed size. 

In this paper, we focus on addressing the size-prescribed $k$-core search problem: Given a graph $G=(V,E)$ and a positive integer $t\leq |V|$, find a size-$t$ subgraph with highest core number, where the core number of a subgraph $H$ as the highest $k$ such that $H$ is a $k$-core subgraph.
% For instance, in social network analysis, it might be desirable to target a $k$-core group of a certain size for maximum impact. Similarly, in protein-protein interaction networks, understanding the characteristics and behaviors of $k$-core groups with a predetermined size can provide valuable insights into the underlying biological processes.

The size-prescribed $k$-core search problem poses unique challenges and calls for innovative methodologies and algorithms. Existing approaches for $k$-core group detection do not directly tackle this specific problem, as they do not consider the size constraint during the search process \cite{barbieri2015efficient,sozio2010community,li2020efficient,cui2014local}. Furthermore, the existing branch-and-cut method, which addresses bounded size subgraphs, is less time-efficient and only suitable for small desired sizes \cite{KaiSizeBounded}. And the existing method \cite{whoShould} relax the $k$-core constraint with a closeness requirement.
Consequently, there is a gap in the literature.

We list two applications of the SPCS problem.
 {\it Event Organization.} Online event recommendation platforms e.g. Meetup, Groupon, Eventbrite commonly employ the social group information to suggest relevant events to users  \cite{luo2014group,boutsis2015personalized}.
When organizing events, size requirements are often specified. Therefore, the goal of platforms is to identify tightly-knit groups of a specific size that can effectively engage participants.
This problem can be formulated as the SPCS problem, where the graph represents individuals and their friendships.
{\it Protein Analysis.} In protein-protein interaction (PPI) networks, where each node represents a protein and each edge represents a protein-protein connection, the proteins within the same $k$-core subgraph often share similar functionalities \cite{wuchty2005peeling,bio_cite1,bio_cite2,bio_cite4}. However, it has been observed that the homophily property does not hold for large $k$-core subgraphs \cite{li2020efficient}. Consequently, there is a need to control the size of the founded $k$-core subgraphs.
The problem of searching for protein clusters of a given size and with the highest core number in the PPI can be modeled as the SPCS problem.

\noindent {\bf Our Contributions.}
The main contributions of this paper can be summarized as follows:

\begin{enumerate}
    \item We formalize the size-prescribed $k$-core search problem. This problem is proven to be NP-hard, W[1]-hard and hard-to-approximate. See Sec.\ref{sec:problem}.
    \item We introduce a novel algorithm called the TSizeKCore algorithm to address the size-prescribed $k$-core search problem. The algorithm incorporates two strategies: a top-down strategy and a bottom-up strategy. 
The top-down strategy starts by locating a large $k$-core within the graph.  Then a size refinement process is applied to reduce the size of the $k$-core until it reaches the desired size. 
On the other hand, the bottom-up strategy begins by identifying a small $k$-core and gradually expands it by iteratively adding nodes until the desired size is achieved. See Sec.\ref{sec:alg}.
\item We validate the effectiveness and efficiency of the proposed algorithms through extensive experiments. In the majority of cases, our algorithm yields optimal solutions. 
% Three detailed case studies are conducted to further analyze the outcomes of our algorithm. See Sec.\ref{sec:exp}.

\end{enumerate}

\section{Related work} \label{sec:related}
The line of work that related to $k$-core subgraph search can be classified into three categories according to the desired sizes of $k$-core: maximal $k$-core subgraph search (e.g., \cite{matula1983smallest, o(m)CoreDecomposition}), minimum $k$-core subgraph search (e.g., \cite{amini2012parameterized,li2020efficient,cui2014local,barbieri2015efficient}) and bouned-size $k$-core subgraph search (e.g., \cite{KaiSizeBounded,whoShould}).
The primary technique to solve the bouned-size $k$-core subgraph search is called branch-and-bound, which is based on the enumeration of all feasible solutions.
However, it should be noted that these exact methods, owing to the fact of numerous candidates, exhibit lower time efficiency and are only applicable to small desired sizes.
To the best of our knowledge, there currently does not exist an efficient algorithm to address the bounded-size $k$-core search problem when the desired size is large. This highlights the need for novel methodologies and algorithms.

\section{The size-prescribed $k$-core search Problem} \label{sec:problem}
Consider a simple connected  graph $G=(V,E)$, where $V$ denotes the set of nodes and $E\subseteq V^2$ denotes the set of edges.
Write $n=|V|$ and $m=|E|$ as the number of nodes and edges in $G$ respectively.
For a subset $S\subseteq V$, write $G[S]$ as the subgraph induced by $S$ in $G$. 
Given a subgraph $H$ of $G$, write $H\cup S$, $H-S$ as the subgraph induced by $V_H\cup S$ and $V_H-S$ in $G$ respectively.
For a node $v\in V$, let $N(v)$ be the nodes that links to $v$ in the graph $G$.

\noindent {\bf $k$-core subgraph \cite{west2001introduction}.} Given an integer $k\in [0,n]$, a subgraph $H$ is a $k$-core subgraph of $G$ if every node in $H$ has at least $k$ neighbors within $H$.

If $H$ is a $k$-core subgraph, then it is also an $i$-core subgraph for all $i=1,2,\cdots, k-1$.

\noindent {\bf Core number.} The core number of a subgraph $H$ is the highest $k$ such that $H$ is a $k$-core subgraph.

Higher core number indicating greater cohesion among graph members. When the core number of $H$ is $|V_H|-1$, then $H$ is a clique. Conversely, when the core number of $H$ is $0$, then $H$ is disconnected.

\smallskip
\noindent {\bf The size-prescribed $k$-core search (SPCS) Problem.} Given a graph $G=(V,E)$ and a positive integer $t\leq |V|$, the SPCS problem searches for a size-$t$ subgraph  with highest core number.

It is easy to see that the $q$-clique problem is a special case of the SPCS problem and thus we have the following results.
% To see this, set $t=q$ and observe that if a $q$-clique exists, then the highest core number of size-$t$ subgraph is exactly $t-1$. Conversely, if the SPCS problem outputs a subgraph with core number $t-1$, then the subgraph is a $q$-clique. Thus, deciding the existence of a clique of size $q$ is equivalent to finding a solution with core number $q-1$ to the SPCS problem on input $\langle G,q\rangle$.
% Given that the $q$-clique problem has been established as $W[1]$-complete \cite{downey1995fixed}, and hard to approximate within $n^{1-\epsilon}$ for any $\epsilon>0$ \cite{hastad1996clique}, we can readily deduce that the SPCS problem in general is a challenging task.

\begin{theorem}\label{thm:hardness}
The SPCS problem is (a) $W[1]$-hard; (b) hard-to-approximate within $n^{1-\epsilon}$ for any $\epsilon>0$. \qed
\end{theorem}

\section{The TSizeKCore Algorithm} \label{sec:alg}
The SPCS problem involves two objectives: maximizing the core number and outputting a subgraph of size $t$. 
Simultaneously satisfying both objectives is computationally challenging, as discussed in Section \ref{sec:problem}.
To address this challenge, it is crucial to establish a priority between these two objectives. Given the potentially large number of $t$-size subgraphs, we firstly maximize the core number and then refine the size of the subgraph to satisfy size requirement. 
% This approach helps reduce the number of candidate subgraphs that need to be considered in subsequent stages, thereby enhancing the overall running time of the algorithm.
We provide a step-by-step description of Alg.\ref{alg:mainalgorithm} below:
\begin{enumerate}
    \item Calculate an upper bound $\overline{k}$ on the core number of all size-$t$ subgraphs. 
    \item Search for a set of $k$-core subgraphs as candidates.
    \item Starting from a $k$-core subgraph, the algorithm adds or removes nodes in an attempt to find a size-$t$ $k$-core subgraph. 
    \item Repeat step 2) and step 3) until it either finds a size-$t$ $k$-core subgraph or failure. If a failure is occurred, decrease $k$ by 1 and go to step 2).
\end{enumerate}
In a connected graph, finding a size-$t$ 1-core subgraph can be achieved by a breadth-first search (BFS) which ensures Alg.\ref{alg:mainalgorithm} will output a feasible solution. 
Notably, Alg.\ref{alg:mainalgorithm} employs two strategies: the top-down strategy and the bottom-up strategy.
The top-down strategy begins with a larger subgraph and iteratively removes nodes until the target size is achieved (Section~\ref{sec:top-down}). Conversely, the bottom-up strategy starts with a smaller subgraph and iteratively adds nodes until the target size is reached (Section~\ref{sec:bottom-up}).
% See a detailed description in Alg.\ref{alg:mainalgorithm}. 

\begin{algorithm}
\DontPrintSemicolon
\KwIn{Graph $G=(V,E)$, integer $t\in [1,n]$}
\KwOut{A core subgraph of size $t$}
\SetKwBlock{Begin}{function}{end function}
Run the algorithm in \cite{matula1983smallest} \;
Let  $\overline{k}$ be the largest core number such that a maximal $k$-core with size $\geq t$ exists\;
Initialize $k\leftarrow \overline{k}$ \;   \label{line:upperbound}
\While{$k> 1$}{
Run  $GetKcore(G,k,t)$ and get $\mathcal{G}$\;  \label{line:getkcore}
\For{each $k$-core subgraph $H\in \mathcal{G}$}{
\uIf{$|V_H|=t$}{
return $H$
}
\Else{
$H'\leftarrow \textit{SizeRefinement(H,t,k)}$ \;  \label{line:size refine}
% $H_1\leftarrow \textit{critical}(H,t,k)$ \;
% $H_2\leftarrow \textit{randomSpeed}(H_1,t,k,H)$ \;
\uIf{$|V_{H'}|=t$}{
return $H'$
}
}
}
$k\leftarrow k-1$ \;
}
\Return{an arbitrary connected subgraph of size $t$}
\caption{TSizeKCore algorithm}\label{alg:mainalgorithm}
\end{algorithm}

\subsection{The Top-down strategy}\label{sec:top-down}
The top-down strategy consists of two parts: identify large $k$-core subgraphs and refine large $k$-core subgraphs.
Initially, the algorithm utilizes the \textit{GetKcore-TD} subprocedure (Line~\ref{line:getkcore}) to obtain a set of $k$-core subgraphs that are larger than the desired size $t$. Subsequently, the strategy employs the \textit{SizeRefinement-TD} function (Line~\ref{line:size refine}) to remove nodes from the given subgraph $H$ until the resulting subgraph is of size $t$. By starting with a larger subgraph and iteratively removing nodes, the strategy aims to find a subgraph that satisfies both the core number and size requirements.

To identify large $k$-core subgraphs, the maximal $k$-core subgraphs naturally serve as large $k$-core subgraph candidates.

\noindent {\bf Maximal $k$-core subgraph.}
A $k$-core subgraph $H$ is a maximal $k$-core subgraph if for any subset $S\subseteq V-V_H$, $H\cup S$ is not a $k$-core subgraph.
Maximal $k$-core subgraphs can be enumerated in $O(m)$ \cite{matula1983smallest}. 

\smallskip
\noindent {\it \bf GetKcore-TD} subprocedure:
Given a graph $G$, a core number $k$, and a size requirement $t$, the subprocedure outputs all maximal $k$-core subgraphs that are larger than $t$.
We directly employ the enumeration algorithm in \cite{matula1983smallest}. 
% For more detailed information on the enumeration algorithm, please refer to \cite{matula1983smallest}.

Once a large $k$-core subgraph is identified, the Top-down strategy begins to reduce its size to achieve the desired size $t$ while maintaining its core number.

\smallskip
\noindent \textit{\bf SizeRefinement-TD} subprocedure takes as input a graph $H$, a core number $k$, and a size requirement $t$, and outputs either a $t$-size $k$-core subgraph or failure.  The procedure iteratively selects a node from $H$. For each selected node $v$, the subprocedure checks whether there exists a $k$-core subgraph of size larger than $t$ in $H-{v}$. If such a subgraph exists, then $H$ is updated with the $k$-core subgraph found in $H-{v}$. The subprocedure returns $H$ if it is of size $t$. The subprocedure continues this process until $H$ reaches the desired size $t$ or all nodes have been checked.  When each node has been checked by the procedure, the resulting subgraph $H$ is a minimal $k$-core subgraph. This means that no subset of nodes can be further removed from $H$ while still maintaining the $k$-core property.
Notice that the order of node removal is determined randomly and thus might result in different outputs.

% However, the \textit{SizeRefinement-TD} procedure is a heuristic and may not find a $t$-size $k$-core subgraph even if one exists in $H$. This is because the sequence of nodes used for removal may not be optimal. Nevertheless, the procedure has been shown to work well in practice and is efficient to implement. Modifications such as trying different sequences of nodes or using other heuristics can be made to improve performance.

\begin{algorithm}
\DontPrintSemicolon
\KwIn{$k$-core graph $H$, integers $k,t<n$}
\KwOut{A $k$-core subgraph of size $t$ or failure}
\SetKwBlock{Begin}{function}{end function}
\setcounter{AlgoLine}{0}
\For{each node $v$ of $H$}{
        $H' \leftarrow H$ \;
       Remove $v$ and its incident edges from  $H'$ \;
        $\mathbb{C} \leftarrow $ all maximal $k$-core subgraphs of $H'$ that has at least $t$ nodes \;
        \uIf{$\exists C\in \mathbb{C}$ such that $|C|=t$}{return $C$}
       \uIf{$\exists C\in \mathbb{C}$ such that $|C|>t$}{$H\leftarrow H'$}
    }
\Return{failure}
\caption{SizeRefinement-TD}
\label{alg:sizeRefineTopdown}
\end{algorithm}
\begin{algorithm}
\DontPrintSemicolon
\KwIn{Graph $G=(V,E)$,  integers $k,t<n$}
\KwOut{A set of $k$-core subgraph of size smaller or equal to $t$ }
\SetKwBlock{Begin}{function}{end function}
\setcounter{AlgoLine}{0}
Initialize an empty set $\mathbb{S}$ \;
$ \mathbb{C} \leftarrow $ all maximal $k$-core subgraphs of $G$ \;
\For{each $H\in \mathcal {G}$}{
         \uIf{$|V_H| \leq t$}{
       add the subgraph $H$ into $\mathbb{S}$ \;
          }
 \uIf{$|V_H|>t$}{
       let $H'$ be the residual subgraph of $H$ after randomly removing $|V_H|-t$ nodes \;
       add all maximal $k$-core subgraphs of $H'$ into  $\mathbb{S}$ \;
    }

}

\Return{ $\mathbb{S}$ }\;
\caption{GetKcore-BU}
\label{alg_GetKcore}
\end{algorithm}

\begin{algorithm}
\DontPrintSemicolon
\KwIn{ Graph $G=(V,E)$, $k$-core subgraph $H=(V_H,E_H)$ of $G$, integers $k,t<n$}
\KwOut{A $k$-core subgraph of size $t$ or failure}
\SetKwBlock{Begin}{function}{end function}
\setcounter{AlgoLine}{0}
\While{$|V_H|<t$ and  $\exists v\in V\setminus V_H$ such that  $N(v)\cap V_H\geq k$ }{
    add $v$ and its incident edges with nodes of $V_H$ into $H$
}
\uIf{$|H| = t$}{
    return $H$ \;
}
\Return{failure}
\caption{SizeRefinement-BU}
\label{alg: SizeRefinement_BU}
\end{algorithm}

\subsection{The Bottom-up strategy}  \label{sec:bottom-up}
The bottom-up strategy is an alternative strategy of the top-down strategy.
The strategy consists of two parts: identify large $k$-core subgraphs and refine large $k$-core subgraphs.
In this strategy, the {\it GetKcore-BU} subprocedure (Line~\ref{line:getkcore}) is called to output a set of $k$-core subgraphs that are smaller than $t$. The {\it SizeRefinement-BU} subprocedure (Line~\ref{line:size refine}) is then used to add nodes from the input graph into the given subgraph $H$, until the resulting subgraph is of size $t$. 
Similarly, the minimal $k$-core subgraphs serve as small $k$-core subgraph candidates.

\noindent {\bf Minimal $k$-core subgraph.}
A $k$-core subgraph $H$ is a minimal $k$-core subgraph if for any subset $S\subseteq V_H$, $H-S$ is not a $k$-core subgraph.

There can exist an exponential number of minimal $k$-core subgraphs making it NP-hard to enumerate them. For example, consider a clique graph with $n$ nodes. For any $k$ in the range $[1, n-1]$, the number of minimal $k$-core subgraphs is ${n \choose k+1}$, which grows exponentially with $n$.
% Furthermore, finding all minimal $k$-core subgraphs is .

The \textit{GetKcore-BU} subprocedure aims to provide $k$-core subgraphs with sizes smaller than $t$. The subprocedure generates subgraphs that are smaller than $t$ but may not be minimal. Such subgraphs can then be further refined through the {\it SizeRefinement-BU} procedure.

% \smallskip
\noindent
{\it \bf GetKcore-BU} subprocedure takes as input a graph $G$, a core number $k$, and a size requirement $t$, and returns a set of randomly selected $k$-core subgraphs that have a size smaller than $t$. This subprocedure operates by first selecting a subgraph of size $t$ at random from the input graph $G$. Then, all maximal $k$-core subgraphs within the selected subgraph are identified and returned. It is important to note that the output of this subprocedure is a result of a random process.
More details on this approach can be found in Alg.\ref{alg_GetKcore}.

Once a small $k$-core subgraph is identified, the Bottom-up strategy begins to add nodes into it to achieve the desired size $t$ while maintaining its core number.

% \smallskip
\noindent 
{\it \bf SizeRefinement-BU}  
Suppose $H$ is a $k$-core subgraph of size smaller than $t$. To find $t$ nodes that induce a $k$-core subgraph, the subprocedure iteratively adds nodes that has at least $k$ neighbors from $G$ into $H$.  If no such node is found, the subprocedure stops and declare a failure.
By the definition of $k$-core, it is clear that the addition does not deteriorate the core number of $H$.
More details can be found in  Alg.\ref{alg: SizeRefinement_BU}.

\section{Experiments} \label{sec:exp}
% \subsection{Experimental Setting}
\noindent {\bf Algorithms:}

 {\it S-greedy} algorithm, originally proposed by Barbieri et al. \cite{barbieri2015efficient}, is a heuristic algorithm designed to solve the minimum $k$-core search problem. In our work, we adapt this algorithm to target a prescribed-size subgraph, addressing the size-prescribed $k$-core search problem.

 {\it TSizeKcore-TD}. The proposed TSizeKCore algorithm with top-down strategy is refereed to as {\it TSizeKcore-TD}. 

 {\it TSizeKcore-BU}. The proposed TSizeKCore algorithm with bottom-up strategy is refereed to as {\it TSizeKcore-BU}.

  {\it Critical}. We also include a naive strategy, the {\it Critical} strategy, as a benchmark. This strategy operates similarly to the top-down strategy. The distinction lies in that it allows repeated selection of nodes and directly checks whether the resulting subgraph $H$ is a $k$-core. It terminates when either a size-$t$ $k$-core subgraph or a critical $k$-core is achieved. A critical $k$-core subgraph is a $k$-core subgraph where for each $v\in V_H$, $H-\{v\}$ is not a $k$-core.

We calculate an upper bound $\overline{k}$ on the optimal core number (see line~\ref{line:upperbound} in Alg.\ref{alg:mainalgorithm}). The closer the core number of the outputted subgraph is to the upper bound, the better the solution is considered to be. In particular, if the outputted subgraph is a $\overline{k}$-core, then it is an optimal solution.
% Note that while a minimal $k$-core must be critical, not all critical $k$-cores are necessarily minimal $k$-cores. 
% We include the critical strategy as a benchmark to further validate the effectiveness of our proposed strategies.

\noindent {\bf Datasets.}
We conducted experiments on ten real-world datasets to evaluate the performance of our algorithm. 
% Seven datasets  {\it Arenas}, {\it Friend}, {\it P2P8}, {\it Lastfm}, {\it Hepph}, {\it Enron}, {\it Gplus}, {\it DBLP}, {\it Youtube}, were chosen to test our algorithm in a social network setting, as it represents human interactions.
% The {\it Yeast} dataset could be suitable for testing our algorithm in a biology setting.
These datasets are collected from SNAP and KONECT.
\begin{table}[h]
\centering
\caption{Dataset statistics}
\resizebox{0.45\textwidth}{!}{
\begin{tabular}{ccccc}
\hline
Dataset & nodes & edges & $avg. deg.$ & $max. deg.$ \\\hline
% moreno & 241 & 1098 & 9.11 & 34 \\\
Arenas & 1,133 & 5,451 & 9.62 & 71 \\\
Friend & 1,858 & 12,534 & 13.49 & 272 \\\
Yeast & 1,870 & 2,277 & 2.43 & 56 \\\
P2P8 & 6,301 & 20,777 & 6.59 & 97 \\\
Lastfm & 7,624 & 27,806 & 7.29 & 216 \\\
Hepph & 34,546 & 421,578 & 24.4 & 846 \\\
Enron & 36,692 & 183,831 & 10.02 & 1,383 \\\
GPlus & 107,614 & 13,673,453 & 254 & 20 \\\
DBLP & 310,297 & 1,186,302 & 4.64 & 340 \\\
Youtube & 1,134,890 & 2,987,624 & 5.26 & 28,754 \\ \hline
\end{tabular}
}
\label{table_dataSet}
\end{table}

\begin{figure*}
  \centering
    \begin{minipage}[t]{0.21\linewidth}
    \includegraphics[width=\linewidth]{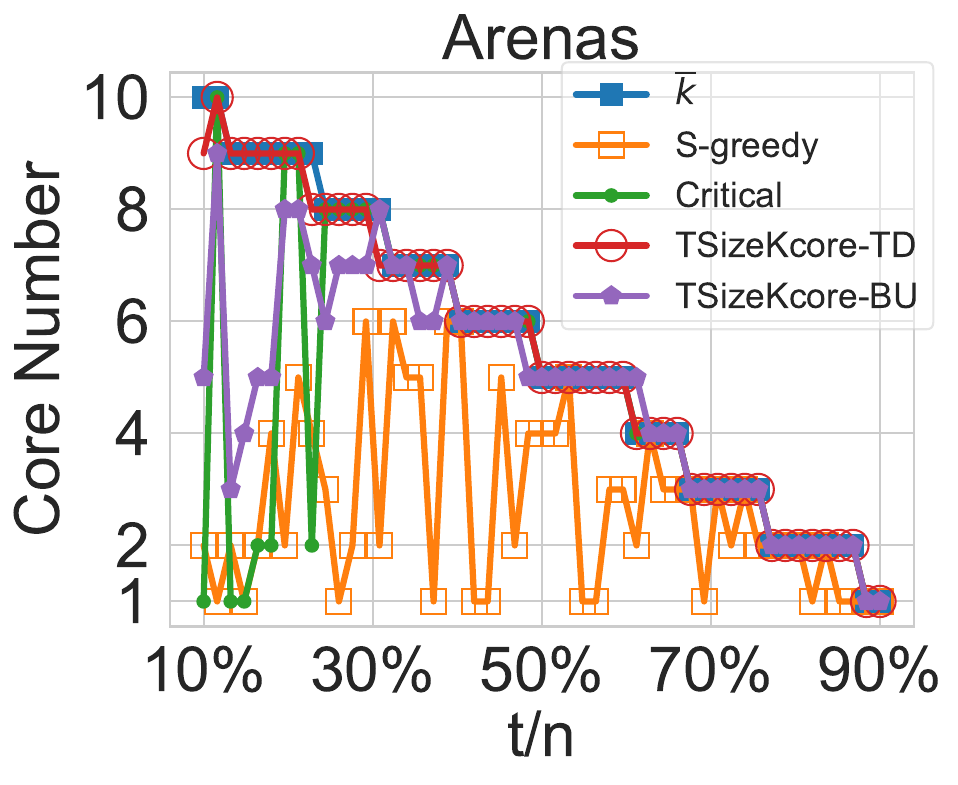}
    \end{minipage}%
    \begin{minipage}[t]{0.2\linewidth}
    \includegraphics[width=\linewidth]{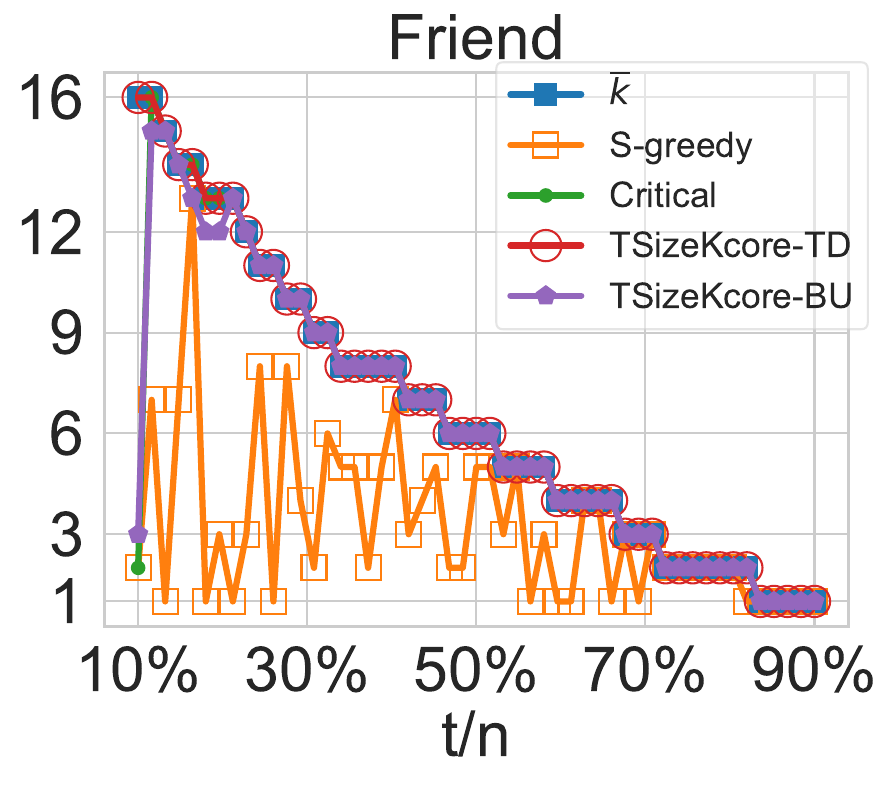}
    \end{minipage}%
    \begin{minipage}[t]{0.193\linewidth}
    \includegraphics[width=\linewidth]{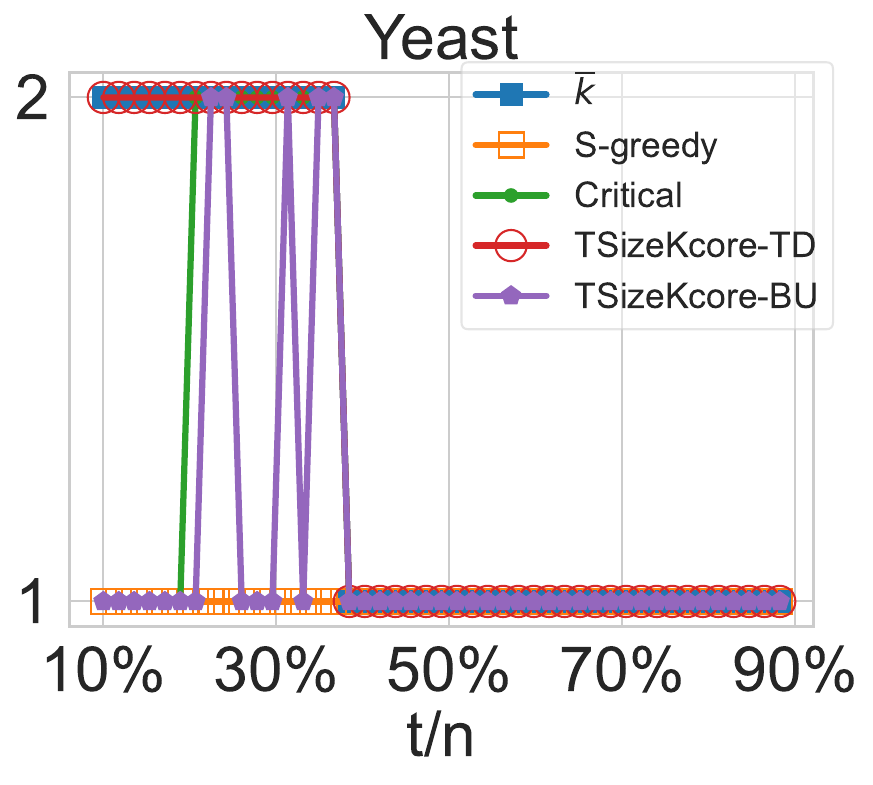}
    \end{minipage}%
    \begin{minipage}[t]{0.2\linewidth}
    % \label{fig:effec_p2p8} 
    \includegraphics[width=\linewidth]{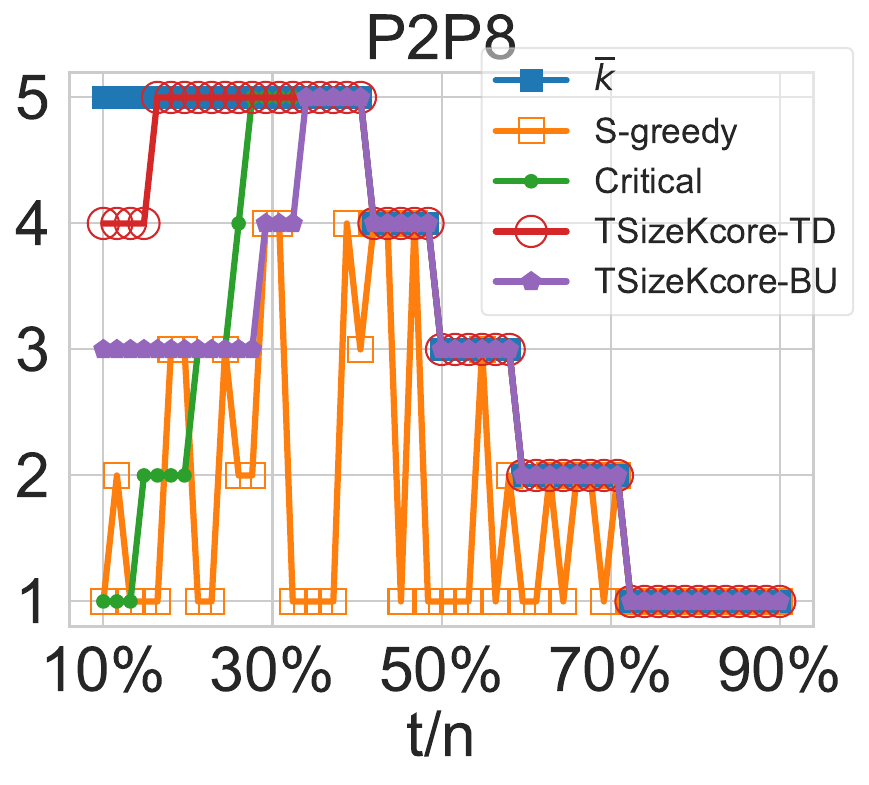}
    \end{minipage}%
    \begin{minipage}[t]{0.2\linewidth}
    % \label{fig:effec_lastfm} 
    \includegraphics[width=\linewidth]{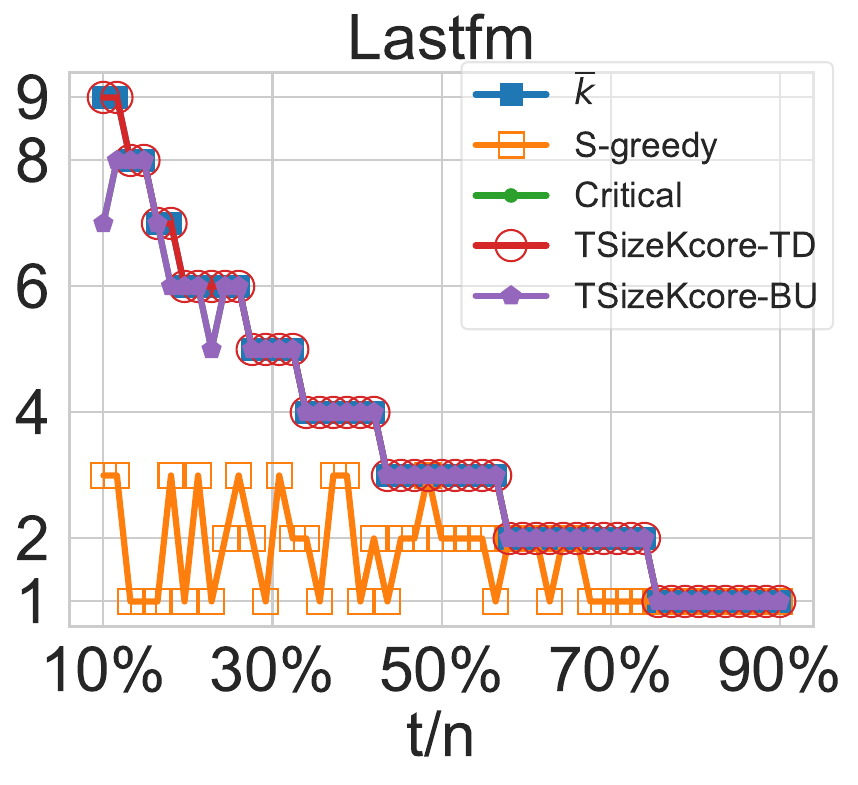}
    \end{minipage}%
    
    \begin{minipage}[t]{0.21\linewidth}
    % \label{fig:medium_hepph_effect} 
    \centering
    \includegraphics[width=\linewidth]{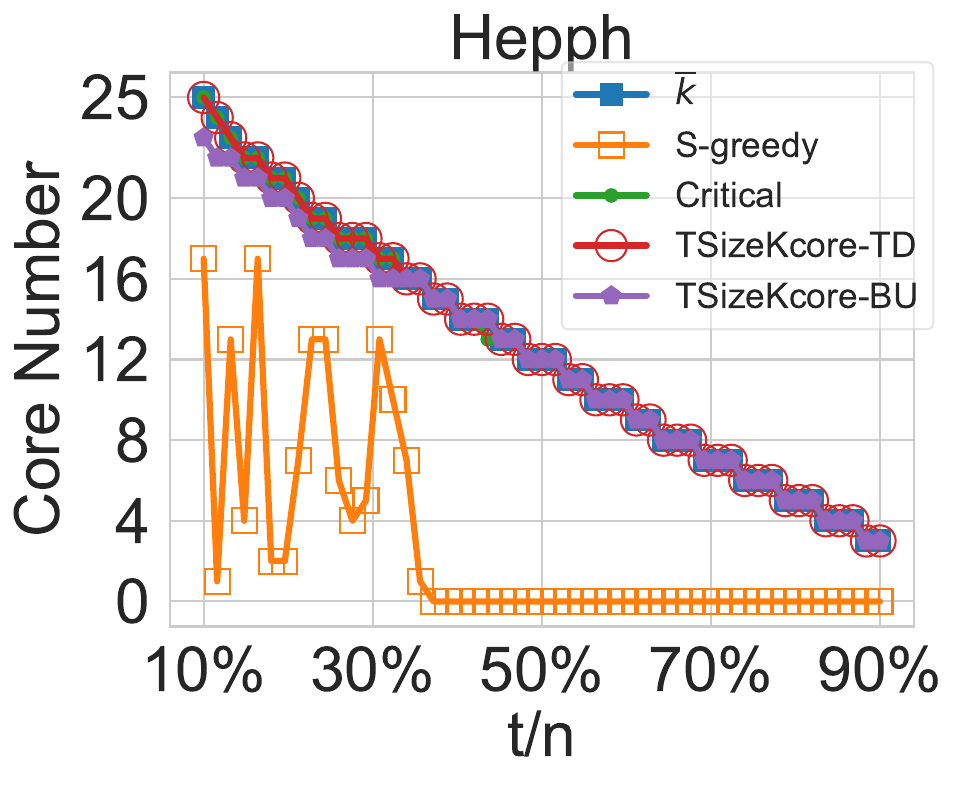}
    \end{minipage}%
    \begin{minipage}[t]{0.2\linewidth}
    % \label{fig:medium_email_effect} 
    \centering
    \includegraphics[width=\linewidth]{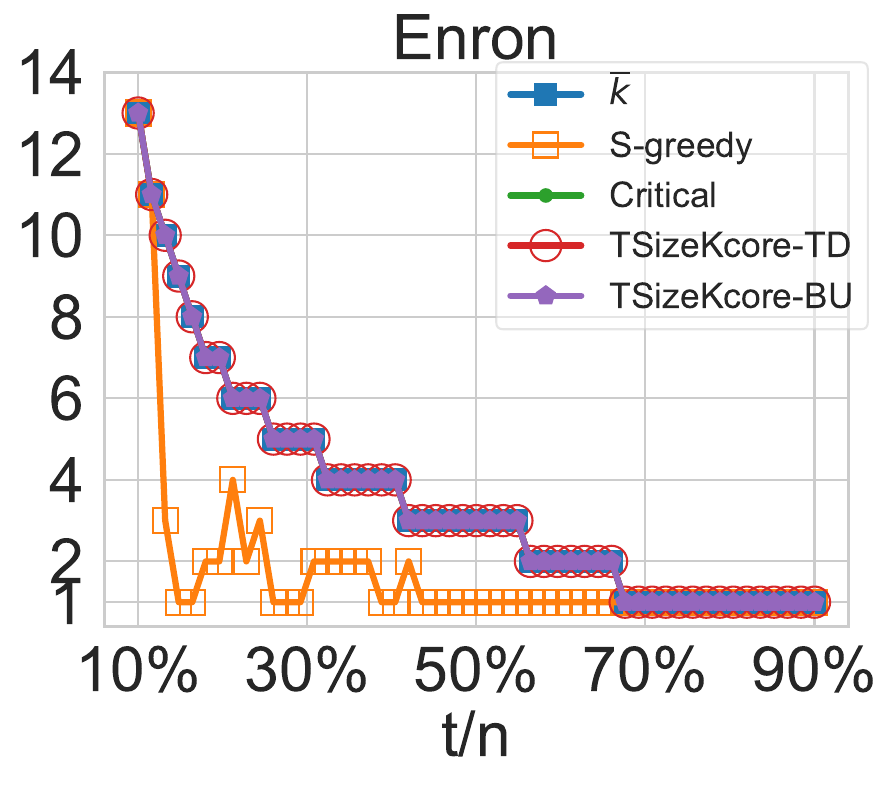}
    \end{minipage}%
    \begin{minipage}[t]{0.21\linewidth}
    \includegraphics[width=\linewidth]{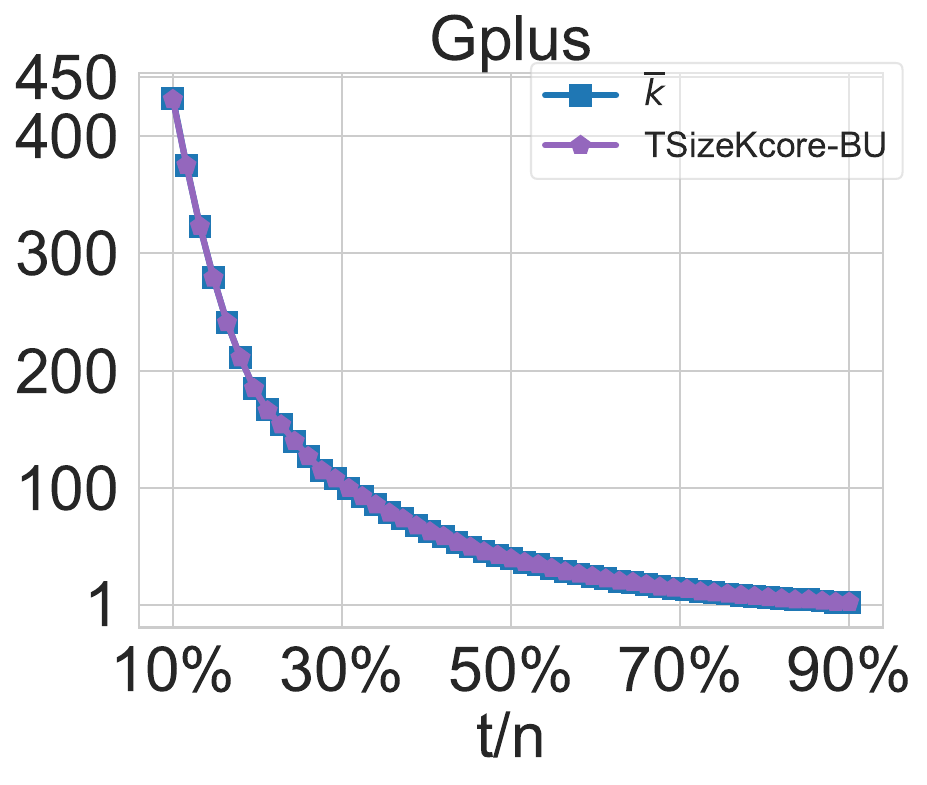}
    \end{minipage}%
    \begin{minipage}[t]{0.2\linewidth}
    \includegraphics[width=\linewidth]{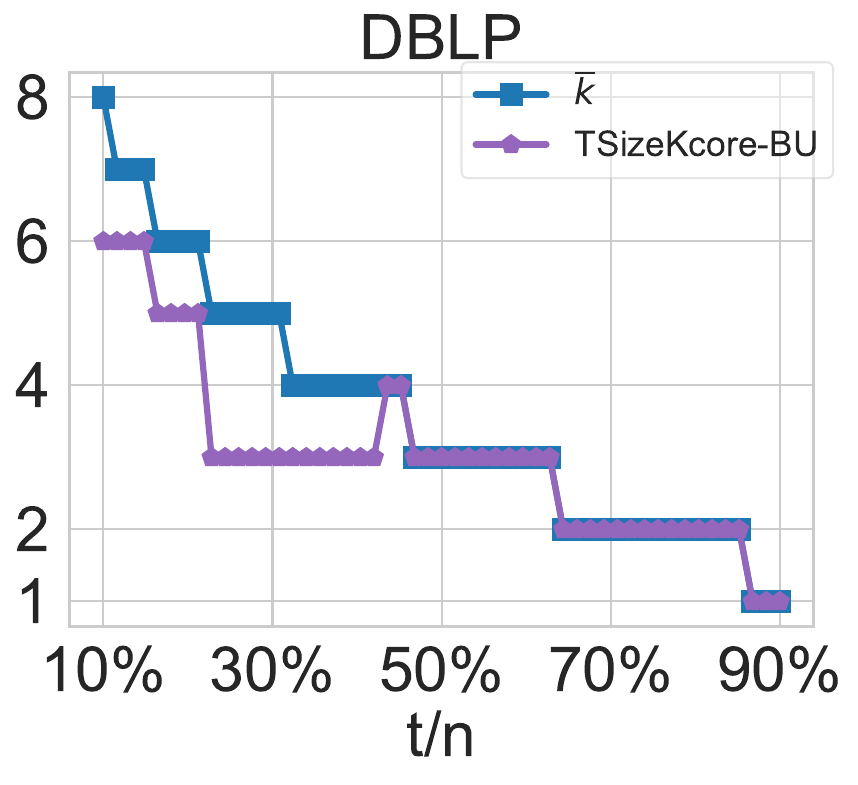}
    \end{minipage}%
    \begin{minipage}[t]{0.2\linewidth}
    \includegraphics[width=\linewidth]{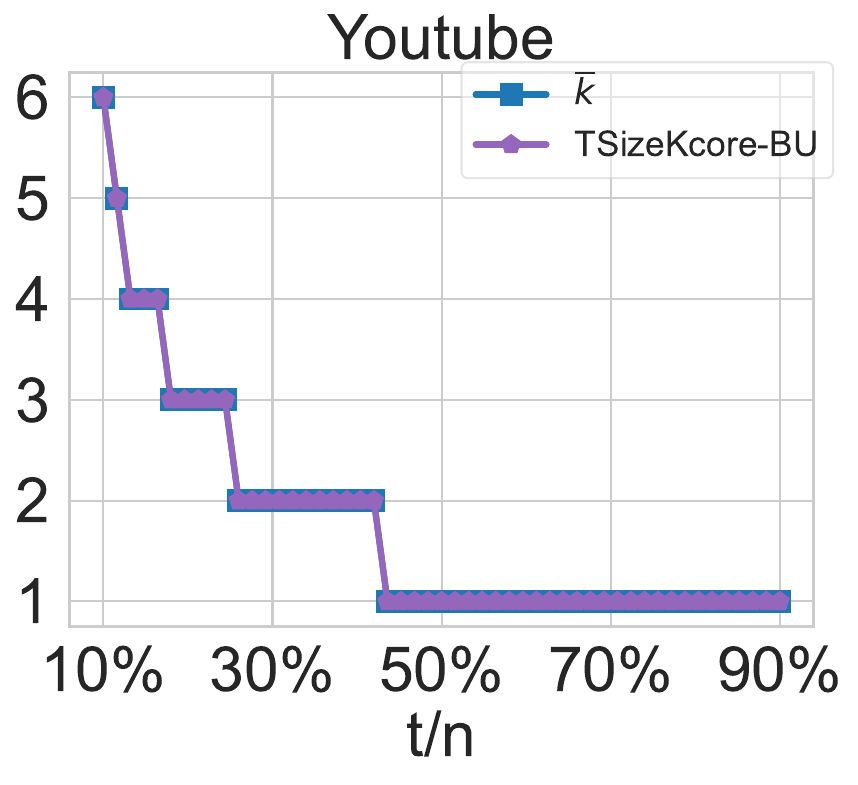}
    \end{minipage}%
  \caption{Effectiveness. The experiment compares the core number of subgraphs outputted by different algorithms with varying $t$.  }
  \label{fig:effective} 
\end{figure*}

\begin{figure*}
    \centering
    \begin{minipage}[t]{0.32\linewidth}
    \includegraphics[width=\linewidth]{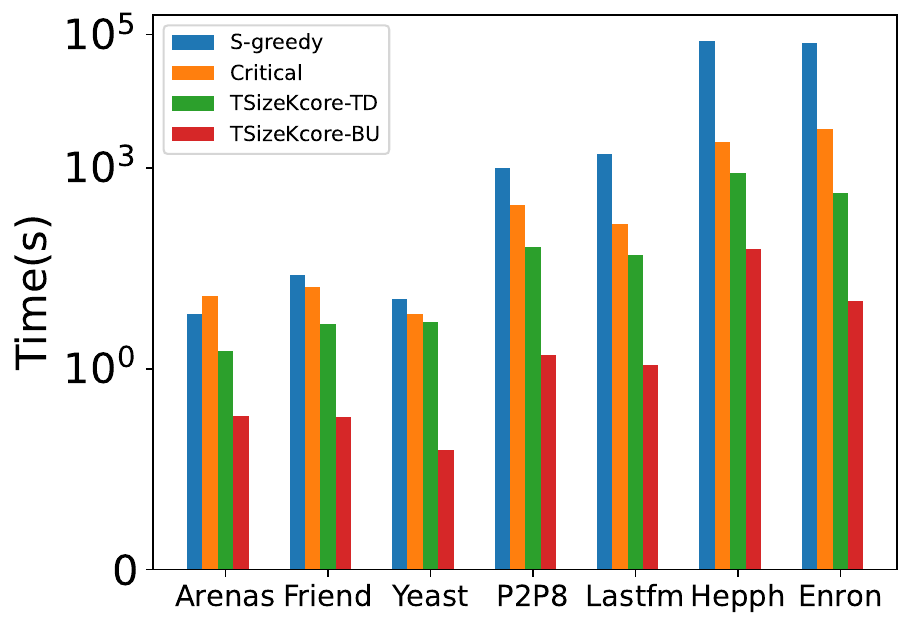}
    \end{minipage}%
    \begin{minipage}[t]{0.24\linewidth}
\includegraphics[width=1\linewidth]{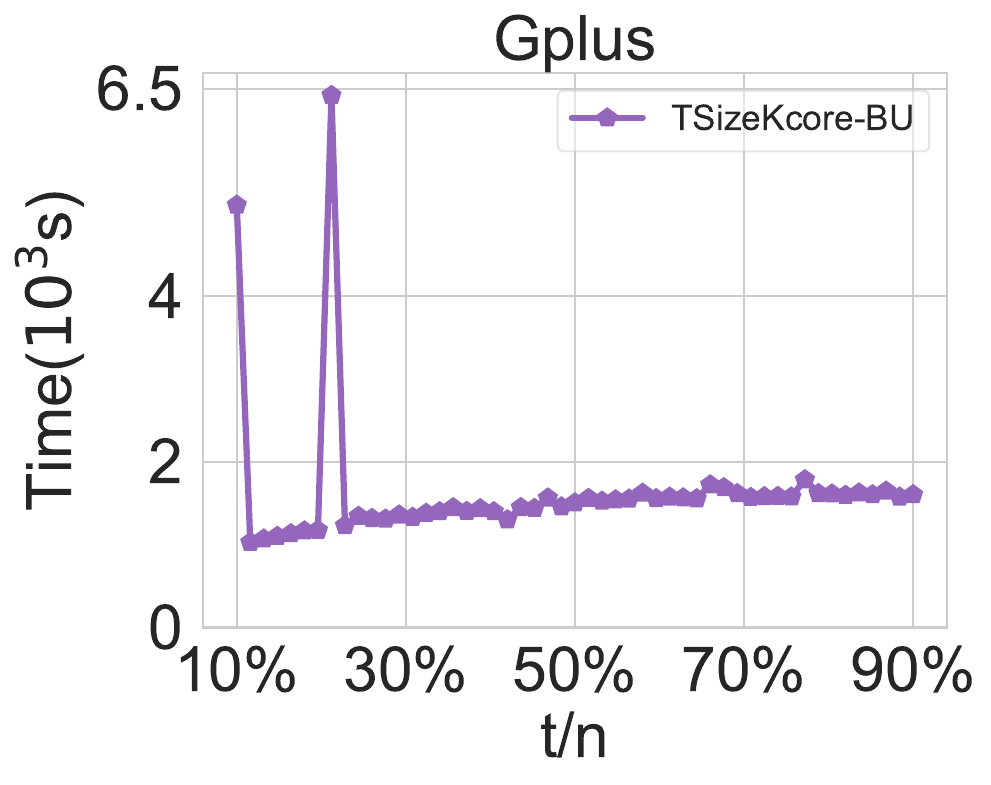}
    \end{minipage}%
    \begin{minipage}[t]{0.21\linewidth}
\includegraphics[width=1\linewidth]{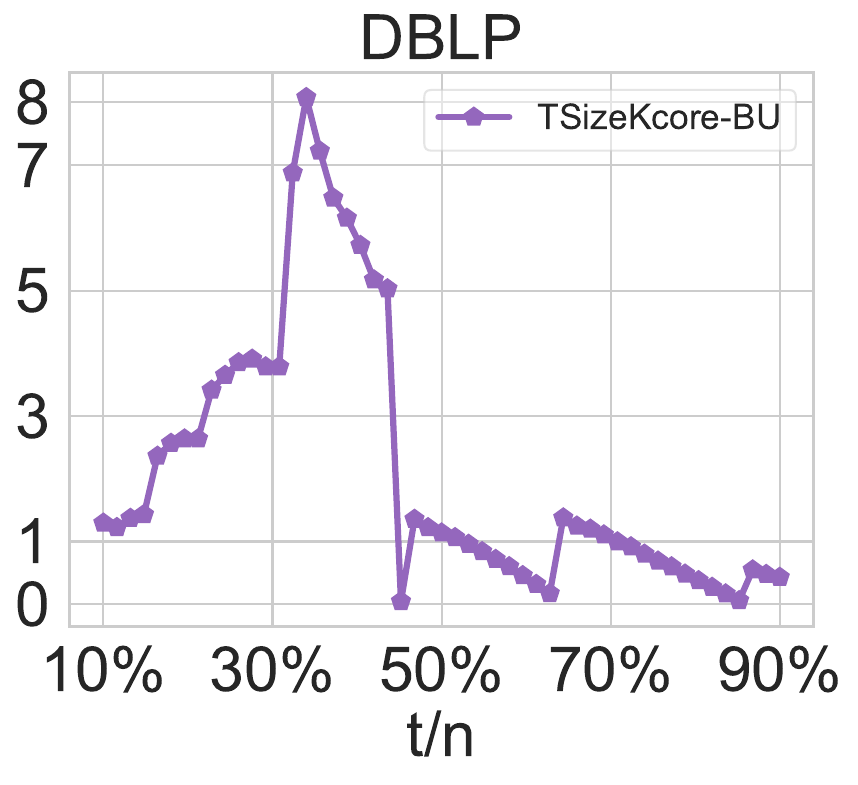}
    \end{minipage}%
    \begin{minipage}[t]{0.22\linewidth}
\includegraphics[width=1\linewidth]{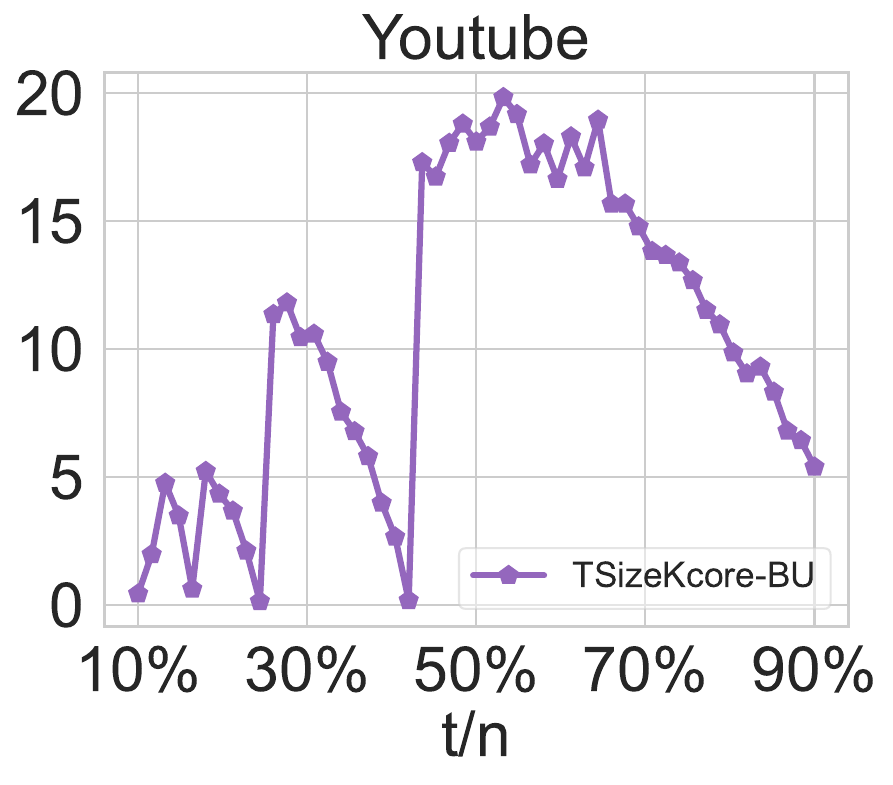}
    \end{minipage}%
    \caption{The average running time of different algorithms. We further list the running time of {\it TSizeKcore-BU} on large datasets. The running time of the {\it S-greedy}, {\it Critical}, and {\it TSizeKcore-TD} algorithms is not listed due to their excessive running time. 
    }
     \label{fig:exp_time} 
\end{figure*}

\noindent 
{\bf Settings.}
% We conducted three experiments: effective experiment, efficiency experiment and an example on event organization.
% The effective experiment compares the core numbers of the outputted subgraphs for different desired sizes $t$ on all ten datasets. The efficiency experiment focuses on comparing the average running time of algorithms.
% We further conduct an experiment in the setting of  event organization. 
All programs are implemented in PYTHON 3.7 and conducted on a machine with Intel Xeon(R) 2.10GHz CPU.
Each result averages 200 outputs.

\noindent 
{\bf Source Code:} \url{https://github.com/turbojob/wrapper}.

% \subsection{Experimental Result}
\noindent 
{\bf Effectiveness experiment.} We compare the core numbers of the outputted subgraphs for different desired sizes $t$ on seven datasets. For large datasets, we conduct solely the {\it TSizeKcore-BU} algorithm due to the high time complexity of other algorithms.
To facilitate comparison on various datasets, we normalize the desired size by the input graph size.
% To ensure statistical significance, each experimental result is an average of 50 outputs.

As shown in Fig.\ref{fig:effective}, the {\it TSizeKcore-TD} algorithm consistently outperforms all other algorithms in all cases. Notably, in most cases, the core numbers outputted by the {\it TSizeKcore-TD} algorithm meet the upper bound $\overline{k}$, indicating that these outputs are optimal solutions.
We also observe a similar performance of the {\it Critical} algorithm and the {\it TSizeKcore-BU} algorithm, despite the {\it TSizeKcore-BU} algorithm having lower time complexity. Furthermore, both algorithms tend to output optimal solutions as the desired size increases, which can be explained by the fact that it becomes less likely for important nodes to be removed, leading to more optimal solutions.
It is remarkable to observe that the {\it TSizeKcore-BU} algorithm consistently outputs near optimal solutions on three large datasets, even when the desired sizes are small and the optimal core number is large. This indicates the effectiveness and accuracy of the algorithm in finding optimal or near-optimal solutions in these scenarios.

\smallskip
\noindent
{\bf Efficiency experiment.}
We focus on the running time comparison of the mentioned algorithms on seven datasets. The running time of other algorithms exceeds 24 hours on three large datasets.
% It is worth noting that the {\it S-greedy}, {\it Classic}, and {\it TSizeKcore-TD} algorithms have high time complexity, making them unsuitable for large datasets. 
We further analyze the average running time of the {\it TSizeKcore-BU} algorithm on the {\it Gplus}, {\it DBLP}, and {\it Youtube} datasets to provide additional insights.

% \begin{figure*}
%   \centering
%     \begin{minipage}[t]{0.25\linewidth}
%     \label{fig:brand_arenas} 
%     \centering
%     \includegraphics[width=\linewidth]{pic/brand/addNode-arenas.pdf}
%     \end{minipage}%
%     \begin{minipage}[t]{0.25\linewidth}
%     \label{fig:brand_friend} 
%     \centering
%     \includegraphics[width=\linewidth]{pic/brand/addNode-friend.pdf}
%     \end{minipage}%
%     \begin{minipage}[t]{0.25\linewidth}
%     \label{fig:brand_yeast} 
%     \centering
%     \includegraphics[width=\linewidth]{pic/brand/addNode-yeast.pdf}
%     \end{minipage}%

%   \caption{Case study: }
%   \label{fig:smallBrand_friend——arenas} 

% \end{figure*}

% \begin{figure*}[ht!]
%   \centering
%     \begin{minipage}[t]{0.34\linewidth}
%     \includegraphics[width=1\linewidth]{pic/application/social/protein 1.png}
%     \end{minipage}%
%     \begin{minipage}[t]{0.34\linewidth}
%     \includegraphics[width=1\linewidth]{pic/application/social/protein 2 delNodes.png}
%     \end{minipage}%
%     \begin{minipage}[t]{0.34\linewidth}
%     \includegraphics[width=1\linewidth]{pic/application/social/protein 3 add nodes.png}
%     \end{minipage}%
  
%   \caption{Illustration of {\it TSizeKcore-BU} algorithm for event organization on {\it Arenas} network. The illustration includes three plots. The plots are arranged from left to right, representing the progression of the algorithm.  
%   }
%   \label{fig:social} 

% \end{figure*}

The experimental results are presented in Fig.\ref{fig:exp_time}. As shown in the first plot of Fig.\ref{fig:exp_time}, the {\it TSizeKcore-BU} algorithm exhibits the lowest running time among all algorithms. Specifically, its average running time remains below 300 seconds on all datasets.
On the other hand, the {\it S-greedy} and {\it Critical} algorithms require longer running times compared to the {\it TSizeKcore-TD} algorithm.
% On large datasets, the running time of the {\it TSizeKcore-BU} algorithm remains below 8,000 seconds on the {\it Gplus} and {\it DBLP} datasets.
For the {\it Youtube} dataset, which contains 1 million nodes and 3 million edges, the running time of the {\it TSizeKcore-BU} algorithm remains below 20,000 seconds. These results further confirm the efficiency of the {\it TSizeKcore-BU} algorithm in handling large-scale datasets. The spikes might result from the large difference between target size and the sizes of existing maximal subgraphs.

In general, the {\it S-greedy} algorithm exhibits the worst performance in terms of both core number and running time.  
The {\it Critical} algorithm shows similar core number to the {\it TSizeKcore-BU} algorithm but has a higher running time. On the other hand, the {\it TSizeKcore-TD} algorithm demonstrates the best performance in terms of core number output, but it requires much running time.

\section{Summary}
% \noindent {\bf Summary}
This paper studies the SPCS Problem which is proven to be computationally hard. 
We propose two algorithms, namely the {\it TSizeKcore-BU} and the {\it TSizeKcore-TD}, which have been shown to provide optimal solutions on most of real-world graphs. 
In particular, the {\it TSizeKcore-TD} algorithm is effective for small datasets where running time is not critical, while the {\it TSizeKcore-BU} algorithm is efficient on large datasets.
% We foresee many ways in which our problem can be extended. For example, the nodes of the graph could have multiple attributes such as genders, ages and locations. By incorporating these attributes, we can extend the SPCS problem with the constraint of desired attributes.  

\noindent 
{\bf Acknowledgement}
This work is supported by China Scholarship Council (CSC) Grant $\#$201906030067.
% \newpage

% \bibliographystyle{ACM-Reference-Format}
% \bibliography{aamas22}
\end{document}